\documentstyle[psfig]{mn}

\let\jnlst=\rm
\def\refjnl#1{{\jnlst#1}}

\def\aj{\refjnl{AJ}}                   
\def\apj{\refjnl{ApJ}}                 
\def\apjs{\refjnl{ApJS}}               
\def\aap{\refjnl{A\&A}}                
\def\mnras{\refjnl{MNRAS}}             

\def\Hgd{H\delta_A+H\gamma_A}
\def\C2{C_24668}

\title[CMRs and Spectral Line Strengths in the Coma Cluster]{Colour-Magnitude Relations and Spectral Line Strengths in
the Coma Cluster}

\author[A. I. Terlevich et al.]
{A.~I.~Terlevich$^1$\thanks{Email: ale@star.sr.bham.ac.uk},~Harald~Kuntschner$^2$, ~R.~G.~Bower$^2$,~N.~Caldwell$^3$,~R.~M.~Sharples$^2$\\
  $^1$University of Birmingham, Edgbaston, Birmingham. B15 2TT.\\
  $^2$University of Durham, South Rd. Durham. DH1 3LE.\\
  $^3$F.L. Whipple Observatory, Smithsonian Institution,PO Box 97,
Amado, AZ 85645 USA}

\date{Submitted 10th March 1999}

\pagerange{\pageref{firstpage}--\pageref{lastpage}} \pubyear{1999}

\begin{document}

\label{firstpage}

\maketitle

\begin{abstract}
  We use the $\rm{C}_24668$, $\rm{Fe}4383$, $\rm{H}\gamma_A$ and
  $\rm{H}\delta_A$ spectral absorption line indices, together with U-
  and V-band photometry of 101 galaxies in the Coma cluster, to
  investigate how mean age and metal abundance correlate with galaxy
  luminosity. In particular, we use the line index measurements to
  address the origin of the colour--magnitude relation (CMR). We
  find that the CMR in Coma is driven primarily by a
  luminosity--metallicity correlation. We additionally show evidence
  for a relation between age and luminosity, in the direction
  predicted by the semi--analytic hierarchical clustering models of
  Kauffmann \& Charlot \shortcite{KauffmannCharlot98}, but this is
  only present in the $\rm{C}_24668$ index models, and could be an
  effect of the lack of non solar abundance ratios in the Worthey
  models used.
  
  By comparing deviations from the CMR, with deviations in absorption
  index from analogous `index--magnitude' relations, we find that
  colour deviations bluewards of the mean relation are strongly
  correlated with the hydrogen Balmer line series absorption. We show
  that the properties of these blue galaxies are consistent with the
  presence of a young stellar population in the galaxies, rather than
  with a reduced metallicity.
\end{abstract}


\section{Introduction}
It has been long understood that the colours of early-type galaxies are
governed primarily by the effects of age and metallicity, which, when
increased, cause the spectral energy distribution (SED) 
to become redder
\cite{Renzini86,Buzzoni92,Buzzoni93,Worthey94,CharlotSilk94}. In his
`$3/2$' law ($\Delta \log(t) \sim {3\over2} \Delta{\rm[Fe/H]}$),
Worthey \shortcite{Worthey94} demonstrated how age and metallicity
have a degenerate effect on galaxy colours. The colour--magnitude
relation (CMR), in which the colours of early--type galaxies become
progressively redder with increasing luminosity, and hence increasing
mass of the galaxy \cite{VisvanathanSandage77} is seen in the cores of
rich clusters, in groups, and even seems to be present in field
ellipticals \cite{LarsonTC80}. Traditionally, the slope seen in the
CMR has been attributed to a mass--metallicity sequence
\cite{Dressler84,Vader86}, with the massive galaxies being more metal
rich, and thus redder, than the less massive ones.  This tendency can
naturally be explained by a supernova--driven wind model
\cite{Larson74,ArimotoYoshii87}, in which more massive galaxies can
retain their supernova ejecta for longer than can smaller galaxies,
thus being able to process a larger fraction of their gas before it is
expelled from the galaxy. Given the degeneracy between the metallicity
and age of a stellar population in its broad band colours, it is also
possible that the CMR is an age driven sequence, with the smaller
galaxies being bluer due to a younger mean age of their stellar
populations. An age dependant CMR however neither preserves its slope,
nor its magnitude range with time
\cite{Kodama97,KodamaArimoto97}. Studies of the CMR in high redshift
clusters find a ridge-line slope comparable to that of the local
clusters; furthermore, there is no sign of a change in the range of
magnitude over which the CMR may be traced
\cite{EllisMORPH97,StanfordED98}. This makes an age dependant CMR in
clusters highly unlikely. Given this metallicity driven interpretation
of the CMR, its low levels of scatter in cluster cores implies that
the galaxies are made up from uniformly old stellar populations
\cite{BLE92II,BKT98}.  Even small variations in the ages of the
galaxies would lead to unacceptable levels of scatter in young stellar
populations, whereas old stellar populations have a much smaller age
dependency in their colours.

On closer inspection however, this picture runs into some problems.  At
high redshifts, the fraction of blue galaxies in many of the clusters
increases \cite{BO78,BO84}. Couch \& Sharples \shortcite{CouchSharples87}
spectroscopically investigated the galaxies in some of these
`Butcher--Oemler' clusters. They split their sample into blue and red
galaxies depending on whether they lie on the cluster CMR or not, and
associate the increased fraction of blue galaxies with either ongoing
or recent star formation. They also found that
11 out of their 73 red galaxies showed enhanced Balmer absorption
lines, indicative of recent bursts of star formation, although it is
possible that these red H$\delta$ strong galaxies can be formed by the
truncation of star formation in spiral galaxies
\cite{BarbaroPoggianti97}.

Spectroscopic studies of early-type galaxies in local rich clusters
(Caldwell et al. \shortcite{Caldwell93}, hereafter C93; Caldwell et
al. \shortcite{Caldwell96}) have shown that these low z clusters also
contain a population of galaxies with abnormal spectra.  They exhibit
enhanced Balmer absorption lines, indicative of recent star formation,
but with too weak an [O{\,\small II}] line to be classified simply as
spiral galaxies.  They note the similarity of some of these spectra with the
`red' H$\delta$-strong galaxies of Couch \& Sharples
\shortcite{CouchSharples87}, and find that of the galaxies associated
with the dynamically separate group of galaxies centered on NGC4839 in
the SW of the Coma cluster \cite{Baier84,Escalera92,CollessDunn96},
about $1/3$ have abnormal spectra.

These spectroscopic studies of local and distant clusters have shown
there to be a population of galaxies which, although they appear
photometrically old, have been forming significant quantities of stars
in their recent past.

In an effort to disentangle age and metallicity effects on stellar
populations, Worthey \shortcite{Worthey94} (also Worthey \& Ottaviani,
1997) developed a series of stellar population spectral synthesis
models, to show the dependence on age and metallicity of spectral
features such as the Balmer lines and various metal lines. In a study
of a magnitude limited sample (M$_B < -17$) of early-type galaxies in
the Fornax cluster, Kuntschner \& Davies
\shortcite{KuntschnerDavies98} used these models to show that, as
predicted by the high--z studies, the elliptical galaxies are indeed
uniformly old, and span a range in metallicities. They also showed
that the majority of the (luminosity weighted) young galaxies in the
cluster were low luminosity lenticular systems. However, earlier
studies of mainly field early-type galaxies \cite{Gonzalez93,Trager97}
show just the opposite, with these galaxies seeming to have a uniform
metallicity and a range of ages, yet still forming a well--defined
CMR. This points to the possibility of an environmental dependence of
the CMR, or maybe two completely different mechanisms operating in
cluster and field environments, conspiring to produce similar CMRs.

In this paper we will look at spectral line indices and deviations
from the CMR in the Coma cluster, using both the Worthey
\shortcite{Worthey94} and Worthey \& Ottaviani \shortcite{WortheyOttaviani97}
spectro-photometric evolution models. We use the high precision
photometry of the Coma cluster presented in Terlevich et~al.,
\shortcite{Terlevich99} (hereafter T99, also Terlevich
\shortcite{Terlevich98}) and the spectra of Coma cluster galaxies from
C93. The selection of galaxies and the spectroscopic and photometric
measurements are outlined in \S2.  Our principle goal is to
investigate whether the CMR in Coma is principally driven by age or
metallicity and to determine the cause of deviations from the CMR in
individual galaxies. Our analysis is presented in \S3, followed by a
summary of our conclusions in \S4.

\section{The Data}

The overall aim of this paper requires us to combine accurate
photometry with precise line-index measurements. The aperture
photometry of T99 covers an area of the Coma cluster that is well
matched to the spectroscopic survey of C93. As a result, we are able
to compare the spectroscopic and broad-band colours of 101 galaxies
drawn from within $50'$ of the cluster center\footnote{An ASCII table
of the galaxy colours and line indices is available on request via
email from A.~I.~Terlevich.}. Below we describe the
selection of the spectroscopic catalogue, the spectroscopic
determination of line strength indices, and the matching photometric
measurements.

\subsection{Sample selection}
C93 selected their sample from the extensive galaxy catalogue of
Godwin, Metcalfe \& Peach \shortcite{GMP83} (hereafter GMP). The
parent catalogue is drawn from a 2.6 deg$^2$ field, centered on the
Coma cluster, and is considered complete to $\rm{B} = 20$,
corresponding to $\rm{M}_B < -14.2$\footnote{We take the recessional
velocity of the Coma cluster to be $\sim6800Kms^{-1}$
\cite{CollessDunn96}, and assume $H_0=100Kms^{-1}~Mpc^{-1}$, which
gives a distance modulus of $(m-M)=34.2$.}  . C93 used two criteria to
select early-type galaxies from the GMP sample. Firstly they fitted
the GMP $(\rm{B},\rm{B}-\rm{V})$
\footnote{C93 converted the GMP (b-r) colours
to (B-V) colours using galaxies in common with the compilation of
\protect\cite{Burstein87}}
colour magnitude relation and selected all
galaxies whose colours lay within $\pm0.15$~mag of the fit. Secondly,
they obtained independent morphologies using KPNO 4m and Palomar Sky
Survey plates.  Using these morphologies, and those of Dressler
\shortcite{Dressler80} they removed morphologically late-type
systems from the list, and added some ($\sim25$) morphologically early-type
galaxies which had failed to be included after the colour cut.
Finally they rejected all galaxies brighter than $\rm{B} = 14.3$, as
high velocity dispersion in these more massive galaxies would make their
spectroscopic analysis of weak absorption lines impossible. We have
further trimmed their sample of 137 galaxies by rejecting the ones
which had the lowest S/N and the emission line galaxies, leaving us with a
total of 101 galaxies.

\begin{table}
\caption{The morphological types for the C93 galaxies
\protect\cite{And96,And97}, split into the three coarse bins of
Spiral/Late type, S0 and Elliptical (see text).}
\centerline{
\begin{tabular}{lc}\hline\hline
Morphological type  & Number of galaxies\\ \hline
No morphology       & 31 \\
Spiral/Late-type    & 12 \\
S0                  & 47 \\
E                   & 11 \\ \hline
All types           &101 \\ \hline
\end{tabular}
}
\label{tab:MorphNumbers}
\end{table}

\begin{figure}
  \centering \centerline{\psfig{figure=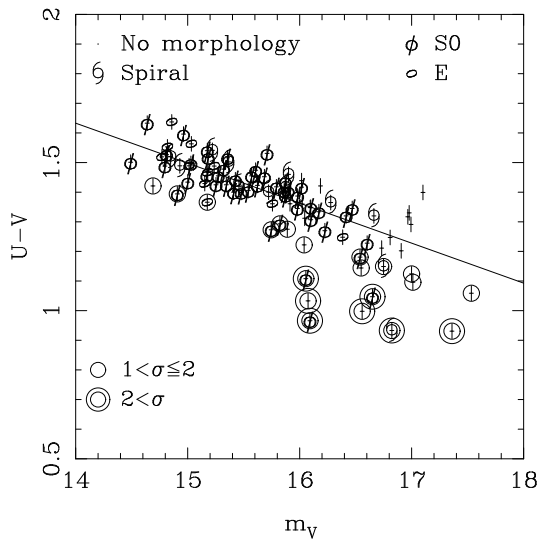,width=7cm,angle=270}}
\caption{The $U-V$ colour-magnitude relation for the subsample of 101 Coma
  galaxies for which we have  appropriate S/N spectra, both colours and
  magnitudes are taken using 8.8~arcsec diameter apertures. The
  standard deviation about the best fit (solid line) is
  $0.07\pm0.01$~mag. The symbols are coded by each galaxy's
  morphological type. Galaxies which lie between 1 and 2 standard
  deviations blueward of the CMR best fit have an additional circle
  drawn around the graph marker. Galaxies which lie more than 2
  standard deviations blueward of the CMR best fit have two extra
  circles drawn about their graph markers.
  The same symbol scheme is used for the galaxies in all the figures
  throughout this paper.}
\label{fig:CM}
\end{figure}

Table~\ref{tab:MorphNumbers} shows the distribution of Andreon et
al. \shortcite{And96} and Andreon et al. \shortcite{And97}
morphological types for the C93 galaxies.  Immediately obvious is the
fact that 12 of the Caldwell early-type galaxies are in fact
late-types according to Andreon et al. \shortcite{And96}. Although
only one of these turns out to lie more than two standard deviations
blueward of the $U-V$~CMR ridge line (see Fig. \ref{fig:CM} and
\S3.1). It must be noted, however, that up to date morphological
information for many of the fainter galaxies in the sample is not
available, and that this is where most of the blue galaxies
reside. Nevertheless, only three out of the seven bluest galaxies have
no determined morphologies.

\begin{figure}
\centering
\centerline{\psfig{figure=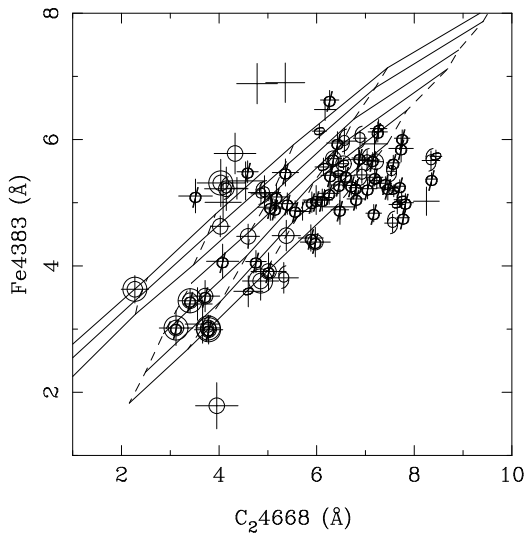,height=7cm,angle=270}}
\caption{The index--index relation between the two metal line indices
used in this paper for the sample of Coma galaxies. A grid of single
age stellar population models from Worthey (1994) is
overplotted. Solid lines connect points of equal age; dashed lines
connect points of equal metal abundance.  The figure shows that for
absorption strengths of $\rm{C}_24668 > 6$~\AA\/, the models do not
match the parameter space occupied by the observations very well. This
is most simply interpreted as an 'overabundance' of $\rm{C}_24668$
compared to Fe. The symbols, which are defined in
Figure~\protect\ref{fig:CM}, are coded by galaxy morphology, and
residual from the CMR.}
\label{fig:C_Fe}
\end{figure}

\subsection{Spectroscopy}
\label{sec:spectroscopy}
The C93 spectra were all taken on the KPNO 4m, using the HYDRA
multi-fibre positioner. They have a spectral resolution of $3.8$~\AA\/
(FWHM), and the fibre aperture was 2~arcsec diameter. In order to compare the
spectra with Worthey stellar population models
\cite{Worthey94,WortheyOttaviani97}, the spectra were re-sampled to the Lick/IDS
resolution by convolution with a Gaussian of wavelength dependant
width. Because the C93 data does not include any observations of stars
in common with the Lick/IDS, the prescription used was that of
Kuntschner \shortcite{Kuntschner98} who used it to correct his $4.1$~\AA\/
(FWHM) resolution Fornax data to the Lick/IDS system. Both the
Kuntschner \shortcite{Kuntschner98} and the C93 spectra cover a similar wavelength
range, and unlike the Lick/IDS spectra, whose spectral resolution
degrades notably towards the blue (see Worthey \& Ottaviani 1997, fig.
7) they have a constant spectral resolution over their entire
wavelength range. The errors on the indices were estimated from the
photon noise of the spectra, and do not take any systematic errors into
account. The most likely source of systematic error is the 
re-sampling of the data onto the Lick/IDS system. 
Nevertheless, the possible systematic errors will have little effect
on the relative comparisons made in this paper. We note that in addition,
our Poisson error evaluation probably underestimates the true error because
response variations from fibre to fibre will introduce extra scatter.

The final step necessary to place the line indices into the Lick/IDS
system is to correct the galaxy spectra for velocity dispersion.
Velocity dispersion corrections for individual line indices were taken
from Kuntschner \shortcite{Kuntschner98}, who calculated them by
artificially broadening stellar spectra (e.g. Davies, Sadler \&
Peletier \shortcite{Davies93}).  The C93 spectra are not of sufficient
resolution and S/N to measure accurate central velocity dispersions
for our galaxies, so instead we construct a `Faber--Jackson' relation
\cite{FaberJackson76} between our V band 8.8~arcsec diameter
aperture magnitudes $m_V$ (see section \ref{sec:photometry}), and the
velocity dispersions ($\sigma$) of Lucey et al.
\shortcite{Lucey97}. A biweight fit (cf., \S\ref{sec:photometry})
gives a relation of \begin{equation} \label{eq:faber_jackson}
log_{10}(\sigma) = (5.6\pm0.6) - (0.22\pm0.01) \times m_V
\end{equation} (with $1\sigma$ bootstrap errors).  We used this
relation to find values of $\sigma$ for those galaxies in the C93
sample which were not present in the Lucey et~al.  \shortcite{Lucey97}
data.

We use the $H\gamma_A$ and $H\delta_A$ Balmer line indices
\cite{WortheyOttaviani97} as the predominantly age sensitive spectral features.
While $H\beta$ is more sensitive to age, it is also more affected by
nebular emission, which can rapidly fill the absorption feature
\cite{Gonzalez93}, so is not used. Higher order Balmer lines are less
sensitive to emission from ionised gas \cite{Osterbrock89}, making an
accurate measurement of the true stellar absorption easier. In order to
increase the signal to noise, we have used $\Hgd$\/ in the final
analysis. The $\rm{C}_24668$ feature is identified by Worthey (1994) as
a particularly sensitive metallicity feature. However it is possible
that it suffers from over abundance problems in the larger, higher
metallicity systems \cite{Kuntschner98}.  Figure \ref{fig:C_Fe} shows
the relation between $\rm{C}_24668$ and $\rm{Fe}4383$ for the Coma
galaxies, overplotted are the theoretical relation from Worthey (1994)
models. Above a value of $\sim6$~\AA, the measured $\rm{C}_24668$ index
values deviate from the region predicted by the models. A similar
effect is seen by Kuntschner \shortcite{Kuntschner98} for early-type galaxies
in the Fornax cluster. However in our data, the $\rm{Fe}4383$ index has
poorer signal to noise than the $\rm{C}_24668$ index, so we will use
both, noting the discrepancy for the high metallicity galaxies.

\subsection{The photometry}
\label{sec:photometry}
It is important to get as good a match as possible between the
photometric and spectroscopic apertures, so that both are measuring the
same part of the galaxy. In the case, for instance, of localised bursts
of star formation occurring in the galaxy nucleus, the large photometric
aperture and the small spectroscopic one, could well both be measuring
different mixes of stellar populations.

In order to match the 2~arcsec diameter of the HYDRA fibres used for
the spectra as best as possible, we use the 8.8~arcsec diameter
photometry from T99. The use of smaller metric apertures was not
possible due to the increased effects of variations in seeing
conditions during the observations. The T99 catalogue covers an area of
$3360\, \rm{arcmin}^2$ to a depth of $V_{13} = 20$~mag (ie., $V$ band
magnitude within a $13''$ diameter aperture) giving Johnson $U$- and $V$-band 
magnitudes for $\sim 1400$ extended objects.  The RMS internal scatter in the
photometry is $0.014$~mag in $V_{13}$, and $0.026$~mag in $U_{13}$ for
$V_{13}<17$~mag.

Throughout this paper we use the biweight minimising technique,
described in T99 (see also Beers, Flynn \& Gebhardt
\shortcite{Beers90}) to fit linear relations to the data. It was
used for its resistance to outlying data points, and its robustness
against non-Gaussian noise distributions.  We derive the residuals of
indices and colours from a mean index or colour - magnitude relation by
the following equation:
\begin{equation}
  \label{eq:residuals}
  \Delta(\rm{X}) = X - (b_{\rm{X}}\times m_V + c_{\rm{X}})
\end{equation}
where X is a line index ($\rm{C}_24886$, $\rm{Fe}4383$ or $\Hgd$), or
($U-V$) colour, $b_X$ and $c_X$ are the slope and intercept of the
best fit relation (see table~\ref{tab:regress}), and $m_V$ is the V
magnitude.

\begin{figure}
\centering
\centerline{\psfig{figure=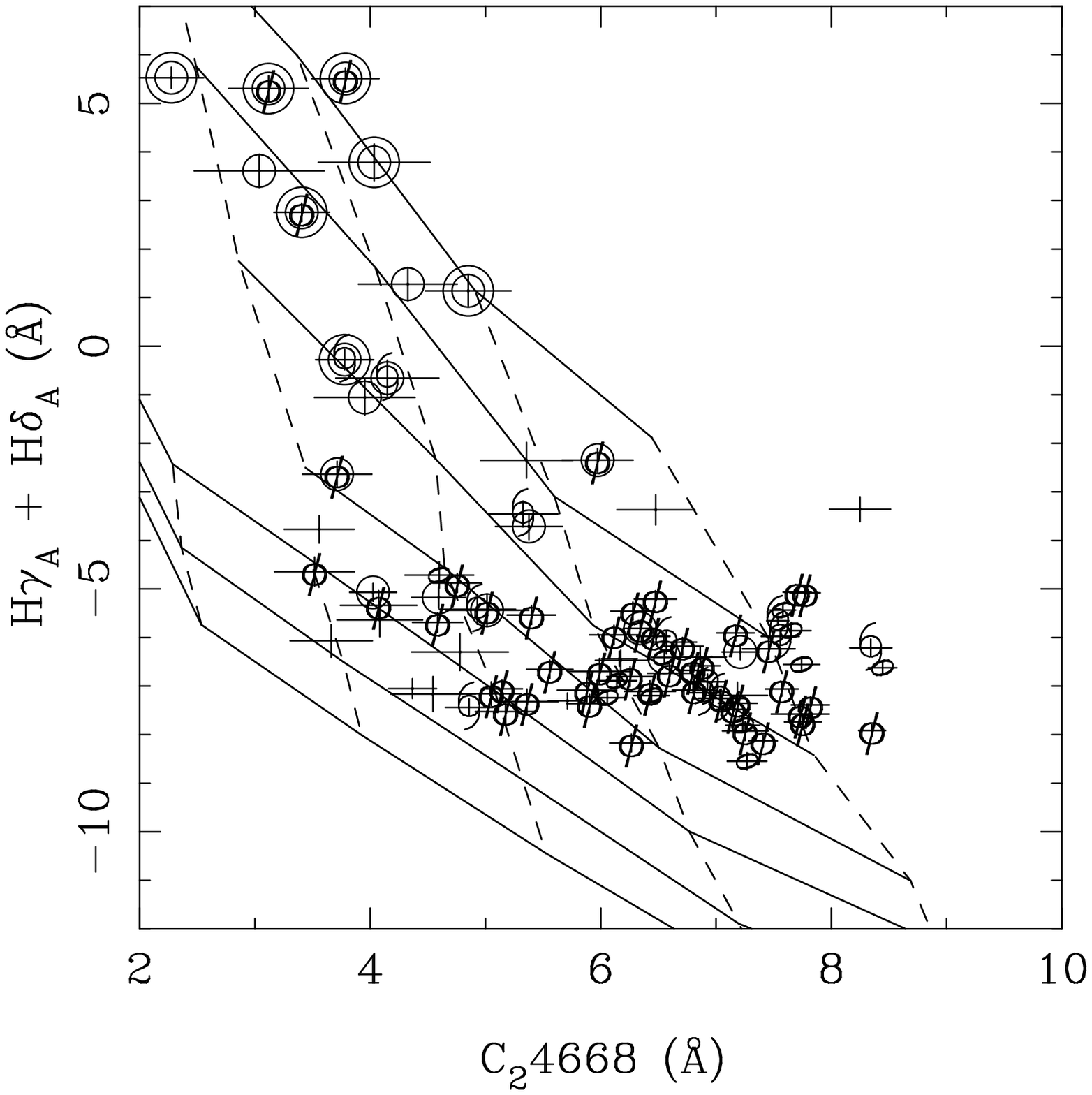,height=7cm,angle=270}}
\caption{ The distribution of galaxies in the $(\rm{C_24668} , \Hgd)$
  plane, with a model grid of single stellar population models
  \protect\cite{Worthey94,WortheyOttaviani97}. Solid lines trace loci of
  constant age (1.5, 2, 3, 5, 8, 12 and 17 Gyrs from positive to
  negative $\Hgd$
  values). Dashed lines trace loci of constant $[Fe/H]$ ( $-0.5$,
  $-0.25$, $0.0$, $0.25$ and $0.5$ from lowest to highest metal line 
  strength). The different symbols
  correspond to each galaxy's morphology, and its offset from the
  colour magnitude relation as defined in figure \protect\ref{fig:CM}.
  }
\label{fig:C_H}
\end{figure}

\begin{figure}
\centering
\centerline{\psfig{figure=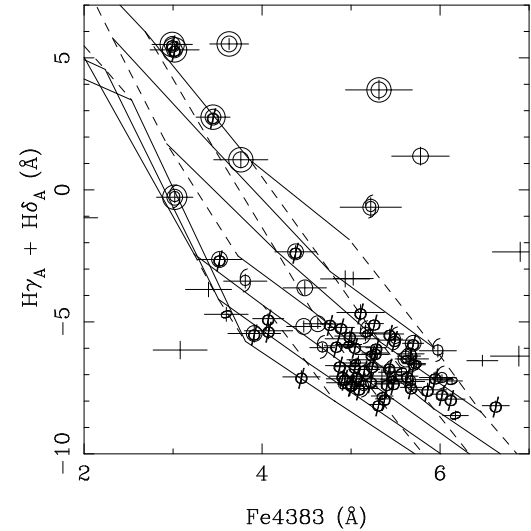,height=7cm,angle=270}}
\caption{ The distribution of galaxies in the $(\rm{Fe}4383 , \Hgd)$
  plane, with a model grid of single stellar population models
  \protect\cite{Worthey94,WortheyOttaviani97}. Solid lines trace loci of
  constant age, from 1.5 to 17 Gyrs. Dashed lines trace loci of
  constant $[Fe/H]$ from $-0.5$ to $0.5$. The different symbols
  correspond to each galaxy's morphology, and its offset from the
  colour magnitude relation as defined in figure \protect\ref{fig:CM}.
  }
\label{fig:Fe_H}
\end{figure}

\section{Implications for star formation histories and the origin of the CMR}
\subsection{The colour magnitude relation}
\label{sec:CMR}
Given the selection criterion of C93, we should expect the CMR to be
prominent in our $(V,U-V)$ data-set. As can be seen in
Figure~\ref{fig:CM}, the bi-weight scatter is indeed small:
$\sigma=0.07\pm0.01$~mag.  Somewhat more surprising is the significant
number of galaxies deviating blueward of the mean relation towards the
faint end. The effect is not seen in the $(B,B-V)$ relation of C93
(see figure 1 in C93), presumably due to the lower sensitivity of
$B-R$ colour to age, combined with their larger photometric errors.
In order to trace the position of these relatively faint and blue
galaxies in the subsequent figures, we adopt a common labelling
convention throughout the paper. The symbols reflect the morphologies
taken from Andreon et~al. \shortcite{And96} and Andreon
et~al. \shortcite{And97}, as defined by the key in figure
\ref{fig:CM}. The number or rings plotted around each symbol indicates
its deviation from the best fit CMR. Points without rings lie either
red-ward of the colour-magnitude relation, or are within $1\sigma$ of
it. Points with only one ring lie between $1\sigma$ and $2\sigma$
blueward of the relation, and points with two rings lie more than
$2\sigma$ blueward of the relation. Unless otherwise stated, when we
refer to `blue' galaxies we are referring to the objects plotted with
{\it at least} one ring, and `red' galaxies are the objects plotted
with no rings.

\subsection{Line Index Diagrams}
\label{sec:CMR_origin}
We have used the $\rm{Fe}4383$, $\rm{C}_24668$ and $\Hgd$\/ indices to
place the galaxies on Worthey's age/metallicity diagnostic diagrams
(Figures \ref{fig:C_H} and \ref{fig:Fe_H}). In these figures, solid
lines trace loci of constant age, from 1.5 to 17 Gyrs (positive to
negative $\Hgd$ values). Dashed lines trace loci of constant $[Fe/H]$
from $-2.0$ to $+0.5$ (lowest to highest metal line strength). Note that
small systematic offsets ($\la0.5$~\AA) between the model and the
observed index systems may well be present. We must stress that these
models represent single stellar population (SSP) systems, whereas our
galaxies contain a mixture of stellar populations of different ages
and metallicities. The indices we measure are therefore a luminosity
weighted mean index of all of the different populations present in
each galaxy, and the higher luminosity of a young stellar population can
mean that even the addition of a small percentage (by mass) of young
stars into an old galaxy, can severely affect it's overall `age'.  For
example, if one adds a $10\%$ (mass), $1{\rm Gyr,[Fe/H]}=+0.5$
population, to an $18{\rm Gyr,[Fe/H]}=0$ population, the resultant
stellar population would have the same line strengths as a $2{\rm
Gyr,[Fe/H]}=+0.25$ SSP \cite{Trager97}. In this paper, when we
refer to a galaxy as `young', we are referring only to its luminosity
weighted mean age.

As we would expect, the blue and red galaxies occupy different areas
of the line index diagrams. The blue galaxies which deviate by more
than $2\sigma$ from the CM relation (Fig.~\ref{fig:CM}) tend to
populate the low age portion of the grid. The trend is common to the
lower colour deviation galaxies as well: only six out of the 20 blue
galaxies lie on the same part of the Worthey grid populated by the red
galaxies.  Both figures imply a young age for the blue galaxies (an
effect driven mainly by the $\Hgd$\/ index); however they seem to
indicate different trends for the the red galaxies which make up the
ridge of the CMR. In Figure~\ref{fig:C_H} red galaxies span a range
both in metallicity and in age, seeming to indicate a far from simple
formation scenario, especially as the more massive galaxies appear to
have the younger ages. We must remember however, that the model grids
at these high metallicities are far from secure. In particular, we
have already noted the overabundance problem in the models for
galaxies with $\rm{C}_24668 > 6$~\AA in \S2.2. If we ignore these
galaxies, the span in age and metallicity of the red galaxies is
reduced.

$\rm{Fe}4383$ does not suffer from the same overabundance effects as
$\rm{C}_24668$, however it also has more of an age dependence,
producing slightly more degenerate model grids. In addition, the
measurements have larger relative errors. In figure~\ref{fig:Fe_H}, we
can see again that the
blue galaxies have younger ages than do the red ones, however they no
longer exclusively populate the low metallicity portion of the grid.
Some of the galaxies are in parts of the plot completely outside the
parameter space covered by the model grid, although these are the
galaxies with the largest error bars in the $\rm{Fe}4383$ data. The red
galaxies have a lower spread in metallicity, and do not follow the same
trend in age as in figure~\ref{fig:C_H}, however, due to the increased
scatter in the $\rm{Fe}4383$ data, and the increased degeneracy in the
model grid, it is still not possible to say that they follow a single
age population model line. This may simply be due to the lower signal
to noise of the $\rm{Fe}4383$ index.

It is worth noting that there is a  morphological
segregation in figures~\ref{fig:C_H} and~\ref{fig:Fe_H}. In both
figures all of the elliptical galaxies populate the old portion of the
model grids, while the S0s and late type galaxies are spread over the
grid. This seems to indicate there are both old and young S0 and late 
type galaxies, but only old ellipticals 
\cite{KuntschnerDavies98,Kuntschner98,Mehlert98}. The overall effect
seems to be connected with galaxy luminosity, with the lower luminosity
galaxies having more varied star formation histories.

\subsection{Colour residuals {\em vs} line strength residuals}
\label{sec:offsets}

\begin{figure}
\centering
\centerline{\psfig{figure=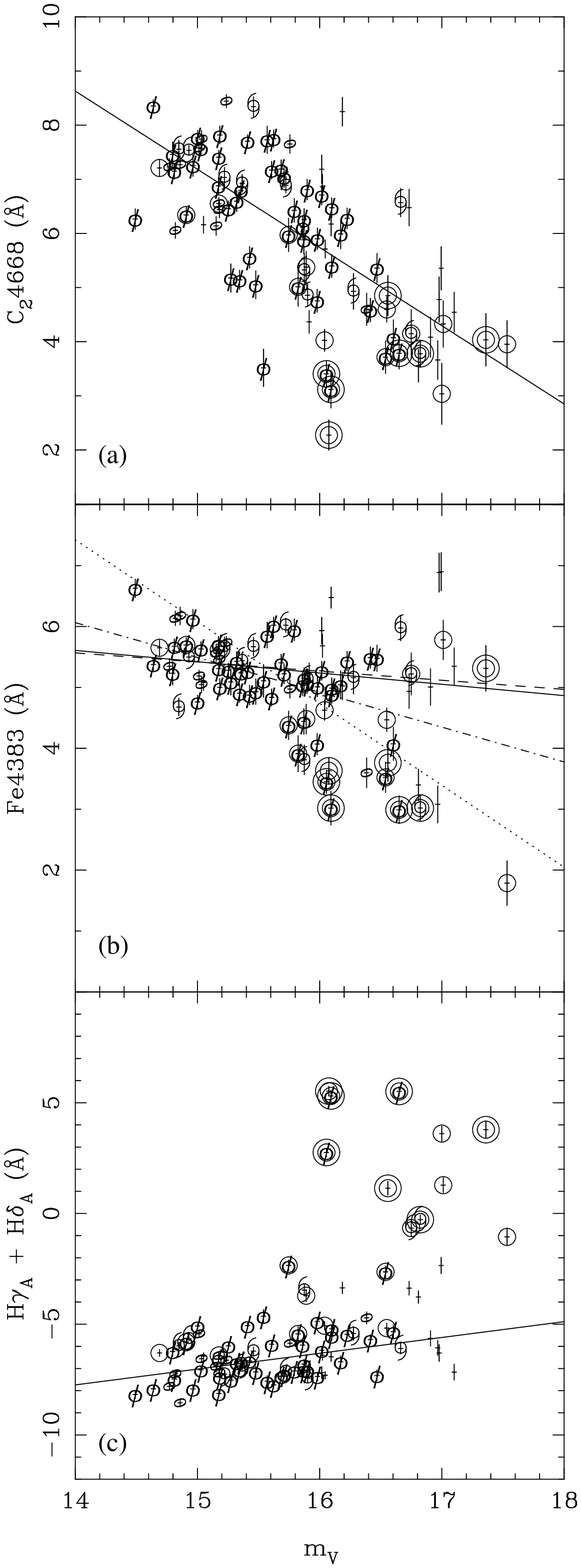,width=7cm}}
\caption{ The relation between the line indices and V magnitude. The
  solid line in each panel shows the biweight regression fit used to calculate
  $\Delta(X)$ (see Table~\protect\ref{tab:regress}). The different
  symbols correspond to each galaxy's morphology, and its offset from
  the colour magnitude relation (see Figure~\protect\ref{fig:CM}). In
  panel b, the dashed line is the biweight scatter minimisation fit to
  only the red galaxies, the dot-dash line shows the biweight scatter
  minimisation fit to galaxies with $m_V < 16$, and the dotted line
  shows the fit to all the data, using an ordinary least squares
  bisector method (see main text).}
\label{fig:corr}

\end{figure}
\begin{figure}
\centering
\centerline{\psfig{figure=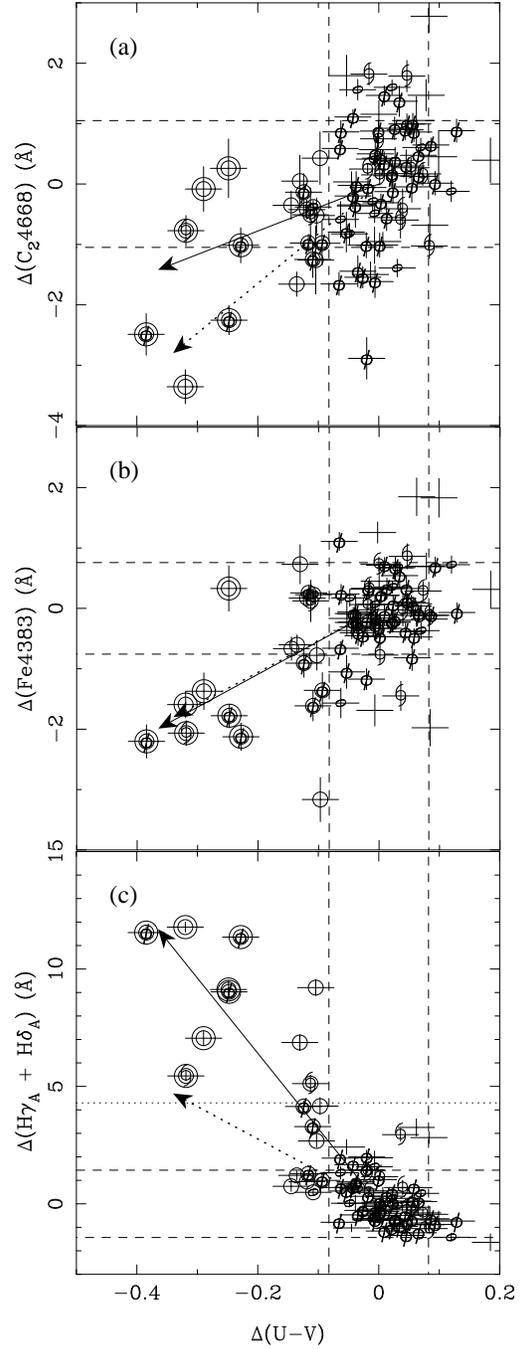,width=7cm}}
\caption{ The deviation of the galaxies from the mean relation in
  the three spectral indices ($\Delta(\rm{C}_24668)$ in panel a,
  $\Delta(\rm{Fe}4383)$ in panel b, $\Delta(\Hgd)$ in panel c), against
  their deviation from the mean relation in colour ($\Delta(U-V)$).
  The fit to the mean relations are calculated from
  Figures~\protect\ref{fig:CM} and~\protect\ref{fig:corr}, and are
  summarised in table~\protect\ref{tab:regress}. The different symbols
  correspond to each galaxy's morphology, and its offset from the
  colour--magnitude relation (see Figure~\protect\ref{fig:CM}). The
  dashed horizontal and vertical lines represent the $\pm1\sigma$
  dispersion in the colours and line strengths. The dotted horizontal
  line in panel c shows the $3\sigma$ dispersion in $\Hgd$. The two
  arrows show how a change in galaxy age (solid arrow) and a change in
  galaxy metallicity (dotted arrow) would affect a 'normal' galaxy. The
  age arrow is calculated by adding a $15\%$ population of 1~Gyr old
  stars to an 8~Gyr old galaxy of the same (solar) metallicity. The
  metallicity arrow shows how an 8~Gyr old solar metallicity galaxy
  would move if its metallicity were changed to $\rm{[Fe/H]}=-0.5$.}
\label{fig:deltas}
\end{figure}

\begin{table*}
\caption{Results of regression analysis with $1\sigma$ bootstrap 
  errors for the line indices and colour {\em vs} V-band magnitude (see    
  figures~\protect\ref{fig:CM} and~\protect\ref{fig:corr}).}
\centerline{
\begin{tabular}{lccc}\hline\hline
Index           & Intercept           & Slope             & Scatter\\\hline
U-V             & $  3.5 \pm 0.7$ mag & $-0.14  \pm 0.01$ & $0.08$ mag \\
$\rm{C_24668}$  & $ 28.9 \pm 3.7$ \AA & $-1.45  \pm 0.06$ \AA$mag^{-1}$ & $1.05$ \AA \\
$\rm{Fe4383}$   & $  7.4 \pm 8.6$ \AA & $-0.14  \pm 0.07$ \AA$mag^{-1}$ & $0.70$ \AA\\
$\Hgd$          & $-18.4 \pm 9.3$ \AA & $ 0.75  \pm 0.12$ \AA$mag^{-1}$ & $1.44$ \AA\\\hline
\end{tabular}
}
\label{tab:regress}
\end{table*}

In section \ref{sec:CMR_origin}, we showed that the blue galaxies which
deviate from the mean CMR have younger luminosity weighted ages than do
the red galaxies. In this section we shall further investigate the
correlation between colour offsets from the CMR, and analogous offsets
from spectral index vs. magnitude relations (IMR)
(Figure~\ref{fig:corr}).
 
We define the mean IMR in the same way as we defined the mean CMR, by
using the biweight estimator (see section~\ref{sec:photometry}) to
get a linear relation between the spectral line strength and
luminosity of a galaxy. We list the results of the fits in
Table~\ref{tab:regress}, and they are shown as the solid lines in
figure~\ref{fig:corr}. The residuals from the mean IMR
are defined using equation~\ref{eq:residuals}.

Figure~\ref{fig:corr} and Table~\ref{tab:regress}, show that we can
define IMRs for all of our spectral indices. The $\rm{C}_24668$
relation (Fig.~\ref{fig:corr}a) shows the best correlation, which fits
in with the interpretation of the CMR as a metallicity sequence, since
$\rm{C}_24668$ is the most metallicity sensitive of our three spectral
indices \cite{Worthey94}. The $\Hgd$\/ relation (Fig.~\ref{fig:corr}c)
also has an obvious IMR. Unlike the $\rm{C}_24668$ IMR, the
red galaxies make up a relation of increasing $\Hgd$\/ with decreasing
luminosity, while most of the blue galaxies are deviant from the
relation.

The $\rm{Fe}4383$ IMR (Fig~\ref{fig:corr}b), like the $\Hgd$ IMR,
shows increased scatter from the blue galaxies. To demonstrate the
effectiveness of the biweight minimisation fit, we show some examples
of different types of fit to the data. The solid line is the biweight
scatter minimisation fit to all the data. The dashed line is the
biweight fit to only the red CMR galaxies, the dot-dash line shows the
biweight fit to galaxies with $m_V < 16$, and the dotted line shows
the fit to all the data, using a method which bisects the ordinary
least squares fits made by minimising the X and the Y residuals
\cite{Isobe90,Fiegelson92}. The biweight fit, and the least squares
fit to only the red galaxies are almost identical. As is the case with
the $\Hgd$\/ relation, most of the deviation comes from the blue
galaxies, so we again take the biweight fit as the definition of the
mean population. It should be noted that there is a much larger
bootstrap uncertainty in the slope of the $\rm{Fe}4383$ IMR than with
any of the others (see Table~\ref{tab:regress}), making the existence
of a non-zero slope to the relation only a $2\sigma$ result.

Out of the three indices we investigate, $\rm{C_24668}$ is the most
sensitive to metallicity, and indeed shows the greatest correlation
with galaxy luminosity. This diagram agrees with the standard paradigm
of a metal abundance driven CMR. The poorer correlation of
$\rm{Fe4383}$ is initially surprising, but seems to reflect the lower
metal abundance sensitivity of the $\rm{Fe4383}$ index relative to
$\rm{C_24668}$, a difference that is enhanced by the so-called `over
abundance' effect (cf., Fig.~\ref{fig:C_Fe}).  This leads to the
interesting possibility that the CMR is in fact driven by a
correlation between luminosity and metal overabundance, however
further investigation of this possibility would require improved
spectra, and is beyond the scope of the present work.  

  Finally, we note that models of the $\rm{Fe4383}$ index have more of
a dependence on age than $\rm{C_24668}$; thus an age gradient along
the CMR (in the sense of fainter galaxies having younger ages) would
tend to steepen the slope. This effect is seen only in the position of
the blue galaxies in Figure~\ref{fig:corr}b.

We now investigate how deviations from the CMR correlate with
deviations from the IMRs. We show the correlations in
Figure~\ref{fig:deltas}. To aid the eye, we show the $\pm1\sigma$
deviations from the CMR and IMRs as dashed horizontal and vertical
lines. The dotted line in panel (c) shows the $+3\sigma$ scatter in
the $\Hgd$ relation. We have used the Worthey \shortcite{Worthey94},
Worthey \& Ottaviani \shortcite{WortheyOttaviani97} models to show how changes
in age (solid arrow) and metallicity (dotted arrow) would affect
galaxies in the CMR/IMR plot. The age arrow was calculated by adding a
15\% (by mass) population of 1Gyr old stars with solar metallicity, to
an 8~Gyr old population with solar metallicity. The metallicity arrow
was calculated by changing the metallicity of an 8~Gyr old population
from $\rm{[Fe/H]}=0$ to $\rm{[Fe/H]}=-0.5$. These arrows can be
thought of as vectors, in that increasing the size of the
post--starburst population, or decreasing the metallicity of the
population further, tends to increase the length of the arrows,
without significantly changing their direction. Note however that if a
galaxy were undergoing a starburst at the time of observation, it
would have Balmer lines (Fig~\ref{fig:deltas}c) in emission, and would
therefore have negative $\Delta(\Hgd)$, but it would still posses blue
colours (negative $\Delta(U-V)$). This emission quickly fades and
turns into absorption, and after $\sim1$Gyr, the galaxies indeed
roughly follow the vector shown in Figure~\ref{fig:deltas}c.

Figure \ref{fig:deltas}a shows deviations from the $\rm{C}_24668$ IMR.
Only three of the $\Delta(U-V) < -2\sigma$ (double--ringed) galaxies
have $\Delta\rm{C}_24668$ values which lie outside the $\pm1\sigma$
range indicated by the dashed lines, with the other four double--ringed
galaxies lying within $\pm1\sigma$ of the $\Delta\rm{C}_24668$
distribution. The age (solid) vector, indicates that we wouldn't expect
much $\rm{C}_24668$ deviation for $\Delta(U-V) = -0.3$, and indeed the
majority of the double--ringed blue galaxies lie close to the age
vector. The three double--ringed galaxies most deviant in
$\rm{C}_24668$, do lie closest to the metallicity vector, however
given the large range in the $\Delta\rm{C}_24668$ indices of the `red'
population, they are not inconsistent with the age vector. 
By contrast, the age and metallicity vectors for the $\Delta\rm{Fe}4383$ 
relation (Figure~\ref{fig:deltas}b) are degenerate, and indeed all but 
a couple of the blue galaxies lie along these vectors.

The correlation between  $\Delta(U-V)$ and $\Delta(\Hgd)$ is the
strongest of the three (Fig~\ref{fig:deltas}c). Like
Figure~\ref{fig:deltas}a, there is a significant separation between
the age and metallicity vectors, and all but one of the double--ringed
blue galaxies lie closer to the age vector than the metallicity
one. This includes the three blue galaxies which lie along the
metallicity vector in Figure~\ref{fig:deltas}a. In this case however, the
range in the $\Delta(\Hgd)$ indices for the `red' galaxies is less
than the equivalent range of $\Delta\rm{C}_24668$ in
Figure~\ref{fig:deltas}a, and cannot be used to account for the
distance of most of the `blue' galaxies from the metallicity vector. 
Additionally, all of the double--ringed blue galaxies deviate from
the mean $\Hgd$ IMR by more than $3\sigma$ (the dotted horizontal
line). These effects can only be accounted for by using a younger mean
stellar population.

\section{Implications}

We have shown using both the $\rm{Fe}4383$ and the $\rm{C}_24668$
indices, that the `red' galaxies which make up the colour--magnitude
relation in the Coma cluster, span a range of metallicities. Whether
this is the sole driving force behind the CMR, or whether there is
also a correlation between age and luminosity is more debatable. There
is no evidence for an age trend with the age-estimates based on the
$\rm{Fe}4383$ index, however the $\rm{C}_24668$ models indicate that
the more luminous galaxies have younger ages. It could be that the
effect is not visible in the $\rm{Fe}4383$ models due to the higher
degree of degeneracy in the age and metallicity tracks for this index,
but it should be noted that any trend in age comes from the ${\rm
C}_24668$ index, at strengths where overabundance effects mean that it
is no longer properly predicted by the Worthey models (see
Fig~\ref{fig:C_Fe} and Vazdekis et~al. \shortcite{Vazdekis96},
\shortcite{Vazdekis97}).  Kuntschner \shortcite{Kuntschner98} corrects
the ${\rm C}_24668$ index in his Fornax cluster galaxies for such
overabundance problems in the models, and shows that much of the age
upturn for the bright ellipticals disappears. If the correlation
between age and luminosity is real however, it would be in the
direction predicted by semi--analytic hierarchical clustering models,
which predict that more luminous cluster galaxies should have younger
luminosity weighed ages, and higher metallicities, than the less
luminous ones \cite{KauffmannCharlot98}. Finally, we note that the
sense of this age gradient is such that the CMR cannot be explained by
an age variation alone. In contrast, even greater metal abundance
variations are required in order to restore the observed slope.

We find that galaxies which deviate bluewards from the
CMR, also deviate from the $(\Hgd)$--magnitude and the
$\rm{Fe}4383$--magnitude relations, but not the
$\rm{C}_24668$--magnitude relation. This is consistent with the colour
deviant galaxies having undergone a secondary burst of star formation
giving young luminosity weighted ages, or showing overall younger ages.
The interpretation of the CMR as a mainly metallicity sequence, with
deviations being caused by age variations seems correct. This
allows the CMR to be used to place constraints on the star formation 
histories of the cluster galaxies, an approached used by many authors
\cite{BLE92II,StanfordED98,EllisMORPH97,BKT98}.

Despite the attempts of C93 to reject late-type galaxies from their
sample, a few still seem to have slipped in. We found that the S0 and
late-type galaxies span a range in ages
\cite{KuntschnerDavies98,Mehlert98}, but all elliptical galaxies are
old. In addition, it is important to note that the younger galaxies
are also lower luminosity galaxies, an effect also seen in the Fornax
sample of Kuntschner \& Davies \shortcite{KuntschnerDavies98}, and in
low luminosity ($M_B>17.5$) Coma S0 galaxies
\cite{Caldwell98}. Worthey \shortcite{Worthey97} also found that lower
luminosity galaxies have a larger spread of ages.  This may indicate
that star formation activity is being `down sized' to lower luminosity
objects as the universe becomes older \cite{Cowie96} or may simply be
an artifact of the greater numbers of lower luminosity galaxies.  Of
the galaxies which lie on the CMR (the red galaxies) we find no
obvious difference between the ages of the early and late types, all
of them have old luminosity--weighted mean stellar ages.  This
includes 10 out of the 12 late-type galaxies in the sample, although
this result should not be too surprising since their spectroscopic
properties will almost certainly be dominated by their bulge.

Finally, we note that the use of CMR--corrected $(U-V)$ colours is an
efficient way of searching for post--starburst galaxies. The $(B,B-V)$
CMR constructed by C93 for these same objects (figure 8 in C93), shows
that them to have a much smaller deviation from the mean population
than our $U-V$ colours do. In fact all our very blue objects, which
deviate blueward by more than two standard deviations from the CMR
colour, have Balmer line strengths that deviate by more than three
standard deviations from the red galaxies. If we were to search for
the PSB galaxies from scratch, and made a colour cut one standard
deviation blueward of the CMR, we would find all of the $3\sigma$
deviant Balmer line galaxies, with only a $50\%$ contamination from
the `normal' population.

\section*{ACKNOWLEDGEMENTS}

We acknowledge the use of STARLINK computing facilities at the
University of Durham and the University of Birmingham. This work was
supported by the PPARC rolling grant for "Extra-Galactic Astronomy and
Cosmology at Durham".

\label{lastpage}
\end{document}